\documentclass{aa}
\usepackage{epsfig}
\usepackage{rotating}
\usepackage{graphicx}
\newcommand{\fermi}{{\sl Fermi}}
\newcommand{\fermin}{{\sl Fermi Gamma-ray Space Telescope}}
\newcommand{\agin}{{\sl Astrorivelatore Gamma ad Immagini LEggero}}
\newcommand{\agi}{{\sl AGILE}}

\newcommand{\gro}{{\sl CGRO}}

\newcommand{\rosn}{{\sl R\"ontgen Satellite}}
\newcommand{\ros}{{\em ROSAT}}
\newcommand{\asca}{{\em ASCA}}
\newcommand{\ascan}{{\em Advanced Satellite for Cosmology and Astrophysics}}
\newcommand{\chan}{{\em Chandra}}
\newcommand{\xmm}{{\em XMM-Newton}}
\newcommand{\vltn}{{\em Very Large Telescope}}
\newcommand{\vlt}{{\em VLT}}
\newcommand{\hstn}{{\em  Hubble  Space Telescope}}
\newcommand{\hst}{{\em HST}}

\newcommand{\fors}{{\em FORS2}}
\newcommand{\forsn}{{\em FOcal Reducer/low dispersion Spectrograph}}

\begin{document}

\title{VLT observations of the two Fermi pulsars PSR\, J1357$-$6429 and PSR\, J1048$-$5832\thanks{Based on observations made with the European Southern Observatory telescopes obtained from the ESO/ST-ECF Science Archive Facility.}}

\author{R. P. Mignani\inst{1,2}
\and
A. Shearer\inst{3}
\and 
A. De Luca\inst{4,5,6}
\and
P. Moran\inst{3}
\and
S. Collins\inst{3}
\and
M. Marelli\inst{5,7}
}

   \institute{Mullard Space Science Laboratory, University College London, Holmbury St. Mary, Dorking, Surrey, RH5 6NT, UK
   \and{Institute of Astronomy, University of Zielona G\'ora, Lubuska 2, 65-265, Zielona G\'ora, Poland}
   \and{Centre for Astronomy, National University of Ireland, Newcastle Road, Galway, Ireland}
   \and{IUSS - Istituto Universitario di Studi Superiori, viale Lungo Ticino Sforza, 56, 27100, Pavia, Italy}
   \and{INAF - Istituto di Astrofisica Spaziale e Fisica Cosmica Milano, via E. Bassini 15, 20133, Milano, Italy}
   \and{INFN - Istituto Nazionale di Fisica Nucleare, sezione di Pavia, via A. Bassi 6, 27100, Pavia, Italy}
   \and{Universit\'a degli Studi dell' Insubria, Via Ravasi 2, 21100, Varese, Italy}
 }

\titlerunning{\vlt\ observations of Fermi pulsars}

\authorrunning{Mignani et al.}
\offprints{R. P. Mignani; rm2@mssl.ucl.ac.uk}

\date{Received ...; accepted ...}

\abstract{Optical observations of pulsars are crucial to study the neutron star properties, from the structure and composition of the interior, to the properties and geometry of the magnetosphere. Historically, X and $\gamma$-ray observations have paved the way to the pulsar optical  identifications.
The launch of the \fermin\ opened new perspectives in the optical-to-$\gamma$-ray studies of neutron stars, with the detection of more than 80 $\gamma$-ray pulsars.
} 
{Here, we aim to search for optical emission from two  \fermi\ pulsars which are interesting targets on the basis of their spin-down age, energetics, and distance. PSR\, J1357$-$6429 is a Vela-like pulsar ($P=166.1$ ms; $\tau =7.31$ kyr), at a distance of $\sim 2.4$ kpc, with a rotational energy loss rate $\dot{E} \sim 3 \times 10^{36}$ erg s$^{-1}$. 
PSR\, J1048$-$5832 is also a Vela-like (P=123.6 ms; $\tau=20.3$ kyr)  pulsar
at a distance of $\sim$ 2.7 kpc and with a  $\dot{E} \sim 2 \times 10^{36}$ erg s$^{-1}$.
The two pulsars and their pulsar wind nebulae (PWNe) are also detected in X-rays by \chan\ and \xmm.  
}   
{No deep optical observations of these two pulsars have been reported so far. We used multi-band optical images ($V,R,I$) taken with the \vltn\ (\vlt) and available in the European Southern Observatory (ESO) archive to search for, or put tight constraints to, their optical emission.}
{We re-assessed the positions of the two pulsars from the analyses of all the available \chan\ observations and the comparison with the published radio coordinates. For PSR\, J1357$-$6429, this yielded a tentative proper motion $\mu = 0\farcs17 \pm 0\farcs055$ yr$^{-1}$  ($70^{\circ} \pm 15^{\circ}$ position angle). We did not detect candidate counterparts to PSR\, J1357$-$6429 and PSR\, J1048$-$5832 down to  $V\sim 27$ and $\sim 27.6$, respectively, although for the former we found a possible evidence for a faint, unresolved object at the \chan\ position.  Our limits imply an efficiency in converting spin-down power into optical luminosity $\la 7\times 10^{-7}$ and $\la 6\times 10^{-6}$, respectively, 
possibly close to that of the Vela pulsar. }
{Observations with the \hstn\ (\hst) are required to identify PSR\, J1357$-$6429 against nearby field stars.
Due to the large extinction ($A_V \sim 5$) and the presence of a molecular cloud complex, near-infrared observations of PSR\, J1048$-$5832 are better suited to spot its candidate counterpart.}

 
   \maketitle

\section{Introduction}
Optical observations of rotation-powered pulsars 
are important to complete the picture of their multi-wavelength phenomenology and play a major role in studying the intrinsic properties of neutron stars,  from the interior structure and composition,  to the atmosphere and magnetosphere properties,  as well as in understanding their formation and evolution  (see, e.g. Mignani\ 2010a,b).  

More than 40 years have gone by since the optical identification of  the Crab pulsar. Since then, only 10 pulsars have been firmly identified in the optical (Mignani\ 2009a; 2011), mostly with the  \hstn\ (\hst) and  the telescopes of the European Southern Observatory (ESO), like the \vltn\ (\vlt), which have played a fundamental role in pulsar optical astronomy  (Mignani\ 2010c; 2009b). Together with those in the radio band, high-energy observations play a pivotal role in paving the way to the optical identification of pulsars.  
In particular,  5 out of the 7 $\gamma$-ray pulsars detected by NASA's {\em Compton Gamma-ray Observatory} (\gro) satellite  (see, e.g.  Thompson\ 2008) have also been detected in the optical, suggesting that $\gamma$-ray detections indicate promising candidates for optical observations, since the emission at both energies seems to correlate with the strength of the magnetic field at the light cylinder (Shearer \& Golden\ 2001; Shearer et al.\ 2010).
The launch of NASA's {\em Fermi Gamma-ray Space} Telescope in June 2008  represents  a revolution in  $\gamma$-ray observations of pulsars.  The {\em Large Area Telescope} ({\em LAT};  Atwood et al. 2009) has detected more than 80 $\gamma$-ray pulsars (Ray \& Saz-Parkinson\ 2010; Caraveo\ 2010).  While in the X-rays systematic observations of {\em Fermi} pulsars are performed (e.g., Marelli et al.\ 2011), for most of them no deep optical observations have been carried out so far (Mignani et al.\ in preparation).  An exploratory survey in the Northern hemisphere has been carried out with 2.5m/4m-class telescopes at the La Palma Observatory but no pulsar has been detected down to $V\approx23$--26 (Shearer et al.\  2011, in preparation), while in the Southern hemisphere a dedicated survey with the \vlt\ is in progress (Mignani et al.\ 2011).  In addition, \vlt\ observations of three {\em Fermi} pulsars, PSR\, J1357$-$6429, PSR\, J1048$-$5832, and the ms binary pulsar PSR\, J0613$-$0200 (Abdo et al.\ 2010)
are available in the public ESO archive. Unfortunately, for the last one an incorrect windowing of the detector cut the pulsar position out of the field--of--view.

PSR\, J1357$-$6429 is a young ($\tau \sim$ 7.3 kyr) Vela-like pulsar discovered in radio (Camilo et al.\ 2004) during the 1347 MHz Parkes multi-beam survey of the galactic plane.  Its period ($P=166.1$ ms) and period derivative ($\dot{P}=3.60 \times 10^{-13}$s~s$^{-1}$) yield a rotational energy loss rate $\dot{E} \sim 3.1 \times 10^{36}$ erg s$^{-1}$ and a magnetic field $B=7.83 \times 10^{12}$ G. The pulsar dispersion measure (DM=$127.2\pm 0.5$ cm$^{-3}$ pc; Camilo et al.\ 2004) yields a distance of $2.4 \pm 0.6$ kpc, according to the NE2001 Galactic electron density models of  Cordes \& Lazio (2002).
X-ray emission at 2--10 keV from PSR\, J1357$-$6429 has been discovered by our group (Esposito et al.\ 2007) from \chan\ and \xmm\ observations. The pulsar X-ray emission appeared point-like, with no obvious evidence of a pulsar wind nebula (PWN).  No pulsations were detected down to a $3 \sigma$ upper limit of 30\% on the pulsed fraction (2--10 keV). However, based on the same \chan\ and \xmm\ data sets Zavlin (2007) claimed evidence of a compact PWN and of X-ray pulsations, 
both of which are now confirmed by new \chan\ and \xmm\ observations (Lemoine-Goumard et al.\ 2011; Chang et al.\  2011). PSR\, J1357$-$6429 has been possibly detected as a $\gamma$-ray pulsar at $\ga 100$ MeV by the \agin\ (\agi) with a detection significance of $4.7 \sigma$ (Pellizzoni et al.\ 2009). Recently, it has been also detected by {\em Fermi} (Lemoine-Goumard et al.\ 2011), while its associated pulsar wind nebula (PWN) has been detected at TeV energies by HESS (Abramowski et al.\ 2011).
PSR\, J1048$-$5832 is also a young (20.3 kyr), Vela-like radio pulsar, discovered during a 1420 MHz radio survey of the galactic plane (Johnston et al.\ 1992). The pulsar period ($P=123.6$ ms) and period derivative ($\dot{P}=9.63 \times 10^{-14}$s~s$^{-1}$) yield a rotational energy loss rate $\dot{E} \sim 2 \times 10^{36}$ erg s$^{-1}$ and a magnetic field $B=3.49 \times 10^{12}$ G.  The DM ($129 \pm 0.1$  cm$^{-3}$ pc; Wang et al.\ 2001) puts the pulsar at a distance of $2.7 \pm 0.35$ kpc.  In the X-rays, PSR\, J1048$-$5832 was observed with the \rosn\ (\ros) at 0.1--2.4 keV (Becker\&Tr\"umper\ 1997) and, soon after, with the \ascan\ (\asca) by Pivovaroff et al.\ (2000), who found possible evidence of extended X-ray emission associated with a PWN.  More recent observations with \chan\ (Gonzalez et al.\ 2006) confirmed the existence of the PWN
although they failed to detect pulsed X-ray emission, with a conservative $3 \sigma$ upper limit of 53\% on the pulsed fraction (0.5-10.0 keV). In $\gamma$-rays, PSR\, J1048$-$5832 was associated with the source 3EG\, J1048$-$5048 (Kaspi et al.\ 2000), detected by the   {\em EGRET}  instrument aboard \gro, both on the basis of a spatial coincidence with the $\gamma$-ray source error box and of the tentative detection of $\gamma$-ray pulsations. Recently, these  were clearly detected at $\ge 0.1$ GeV by {\em Fermi}  (Abdo et al.\ 2009), with a double-peaked light-curve.

Here, we report on the results of an archival \vlt\ survey for {\em Fermi} pulsars.  This paper  is organised as follows: observations,  data reduction and analysis are  described in  Sect. 2, while  results are  presented and discussed in Sect. 3 and 4, respectively.  Conclusions follow.

\begin{table*}[t]
\begin{center}
  \caption{Summary of the available pulsar observations, reporting the exposure times in s (T), the number of exposures (N), the average airmass sec(z) during the sequence of N exposures, the associated average image quality (IQ), and its rms (in parentheses).}
\begin{tabular}{llllllc} \\ \hline
Pulsar   &  Date  &  Filter & T & N & sec(z) & IQ \\  
              & yyyy-mm-dd & & (s) &  & & ($\arcsec$)  \\ \hline
PSR\, J1357$-$6429 & 2009-04-04 	 & $v_{\rm HIGH}$ 	 & 580 	 & 5 	 & 1.35 	 & 0.62 	 (0.03)  \\   
                   		& 2009-04-04 	 & $R_{\rm SPEC}$ & 580 	 & 5 	 & 1.44 	 & 0.58 	 (0.03)  \\   
                   		& 2009-04-22 	 & $v_{\rm HIGH}$ 	 & 590 	 & 20 & 1.35 	 & 0.65 	 (0.03)  \\   
                   		& 2009-04-24 	 & $R_{\rm SPEC}$ & 590 	 & 10 & 1.35 	 & 0.59 	 (0.04)  \\   
                   		& 2009-04-25 	 & $I_{\rm BESS}$ 	 & 200 	 & 8 	 & 1.31 	 & 0.57 	 (0.04)  \\   \hline
PSR\, J1048$-$5832 & 2010-01-11 	 & $v_{\rm HIGH}$ 	 & 750 	 & 4 	 & 1.21 	 & 0.69 	 (0.04) \\  
                                  & 2010-01-13 	 & $v_{\rm HIGH}$ 	 & 750 	 & 8 	 & 1.23 	 & 0.56 	 (0.03)  \\   
                   		& 2010-01-23 	 & $v_{\rm HIGH}$ 	 & 750 	 & 4 	 & 1.21 	 & 0.79 	 (0.04) \\    
                   		& 2010-01-24 	 & $v_{\rm HIGH}$ 	 & 750 	 & 8 	 & 1.21 	 & 0.60 	 (0.03)  \\   
                   		& 2010-02-10 	 & $v_{\rm HIGH}$ 	 & 750 	 & 8 	 & 1.22 	 & 0.65 	 (0.03)  \\   \hline
\label{obs}
\end{tabular}
\end{center}
\end{table*}

\section{Observations and data reduction}

\subsection{Observation description}

Optical images of the PSR\, J1357$-$6429  and PSR\, J1048$-$5832 fields were obtained  with the \vlt\ Antu telescope at the ESO Paranal  observatory between April 2009 and February 2010  (see Tab. 1 for a summary of the observations) and are available in the public ESO archive\footnote{www.eso.org/archive}.   Observations were  performed in service  mode with the \forsn\  (\fors; Appenzeller  et  al.\ 1998), a multi-mode camera for  imaging and long-slit/multi-object spectroscopy (MOS).  \fors\  was equipped with its red-sensitive MIT detector, a mosaic of two 2k$\times$4k CCDs optimised  for wavelengths  longer  than 6000  \AA.   In its  standard resolution  mode,   the  detector  has  a  pixel   size  of  0\farcs25 (2$\times$2 binning) which  corresponds to a projected field--of--view of 8$\farcm3  \times 8\farcm3$  over the CCD  mosaic. However,  due to vignetting, the effective sky coverage of the two detectors is smaller than the projected detector field--of--view and it is larger for the upper CCD chip.   Observations were performed with the standard low gain, fast read-out mode and in high-resolution mode  (0\farcs125/pixel) for PSR\, J1357$-$6429 and in standard resolution mode (0\farcs25/pixel) for PSR\, J1048$-$5832.  In both cases, the target was positioned in the upper CCD  chip. For PSR\, J1357$-$6429,  bright stars close to the pulsar position have been masked using the \fors\ MOS slitlets as occulting bars.  Different filters were used:  $v_{\rm HIGH}$ ($\lambda=5570$ \AA;  $\Delta \lambda=1235$\AA), $R_{\rm SPEC}$  ($\lambda=6550$ \AA;  $\Delta \lambda=1650$\AA), and  $I_{\rm BESS}$ ($\lambda=7680$ \AA;  $\Delta \lambda=1380$\AA).  To allow for  cosmic ray removal and minimise saturation of bright stars in the field, sequences of short exposures (from 200 to 750 s) were obtained per each target and per each filter. The total integration time was 14700 s ($v_{\rm HIGH}$), 8800 s ($R_{\rm SPEC}$), and 1600 s ($I_{\rm BESS}$) for  PSR\, J1357$-$6429  and of 24000 s  ($v_{\rm HIGH}$) for PSR\, J1048$-$5832. Exposures  were taken  in  dark time  and under  photometric conditions\footnote{http://archive.eso.org/asm/ambient-server},
 with an airmass  mostly below 1.3 and sub-arcsecond image quality, as measured directly on the images by fitting the full-width at half  maximum (FWHM) of unsaturated field stars.

\subsection{Data reduction and astrometry}

We reduced the data through  standard packages in  {\em IRAF}  for  bias   subtraction,  and  flat--field   correction using the closest--in--time bias and twilight flat--fields frames available in the ESO archive. Per each band, we aligned and average-stacked the reduced  science images using the  {\em IRAF} task {\tt  drizzle} applying a $3  \sigma$ filter on  the single pixel average  to filter  out residual  hot and  cold pixels and cosmic ray hits.  Since all exposures have been taken with sub-arcsec image quality, we did not apply any selection prior to the image stacking.  We applied the  photometric calibration by using  the  extinction-corrected night  zero points  computed by  the  \fors\ pipeline  and  available through  the instrument  data quality  control database\footnote{www.eso.org/qc}.  
To register the pulsar positions on the \fors\ frames as precisely as possible, we re-computed their  astrometric solution which is, by default,  based on the coordinates of the guide star used for the telescope pointing.  Since most stars from the  Guide Star Catalogue  2 (GSC-2;  Lasker et al.\ 2008) are saturated in the stacked images, we used shorter exposures (10--15 s) of the fields taken with the same instrument configurations as those in Tab. 1 and available in the \vlt\ archive.  We  measured the star centroids through Gaussian fitting  using the Graphical  Astronomy  and   Image  Analysis  (GAIA)  tool\footnote{See star-www.dur.ac.uk/$\sim$pdraper/gaia/gaia.html}  and used  the code  {\tt  ASTROM}\footnote{www.starlink.rl.ac.uk/star/docs/sun5.htx/sun5.html} to compute  the pixel-to-sky coordinate  transformation through an high-order polynomial, which accounts for the CCD distortions.   For both pulsar fields, the  rms  of the astrometric  fits was $\sigma_{\rm r} \sim 0\farcs1$ in  the radial direction.   To this  value  we added  in  quadrature the  uncertainty $\sigma_{\rm tr}=0\farcs1$ of the registration of the \fors\ image on the  GSC2  reference frames, $\sigma_{\rm tr}= \sqrt(3/N_{s}) \sigma_{\rm  GSC2}$,  where  $\sigma_{\rm  GSC2}=0\farcs3$ is  the  mean positional error  of the GSC2  coordinates  and $N_{s}$ is  the number of stars used  to compute the astrometric solution (Lattanzi et al.\ 1997).  After accounting for the $\sim 0\farcs15$ accuracy of the link of the GSC2 coordinates to the International Celestial Reference Frame (ICRF), we thus estimate that  the overall ($1 \sigma$) uncertainty  of our \fors\  astrometry is $\delta_{r}\sim0\farcs2$. 

\subsection{The problem of position: radio vs. X-rays}

As  a  reference for our astrometry, we  started from  the  most recently published radio positions of the two pulsars. Moreover, they both have precise X-ray positions obtained with \chan. The published radio and X-ray coordinates of  PSR\, J1357$-$6429 and PSR\, J1048$-$5832 are summarised in Table 2.

We note that the \chan\ position of PSR\, J1357$-$6429 (Zavlin 2007) is  quite different from that derived from the radio-interferometry observations performed with the Australia Telescope Compact Array (ATCA) by Camilo et al.\  (2004).  Since this apparent inconsistency hampers the correct pulsar localisation on the \fors\ images,  hence the search for its optical counterpart, we independently checked all the available sets of coordinates.  
For PSR\, J1357$-$6429, multiple \chan\ observations are available. Thus, we could use them to check the internal consistency of the \chan\  astrometry. Firstly, we re-computed the X-ray position from the analysis of the two, independent \chan/HRC observations (Esposito et al.\ 2007; Zavlin\ 2007) and we found values consistent with that published in Zavlin\ (2007). The pulsar has been recently observed also with the ACIS instrument (OBS-ID=10880; MJD=55112.28). We downloaded the data from the public \chan\ archive and we found that the pulsar coordinates are  also consistent, within the  nominal \chan\ position uncertainty of $0\farcs6$\footnote{http://asc.harvard.edu/cal/ASPECT/celmon} (90\% confidence level), with those obtained with the HRC. We then averaged the three sets of coordinates, which  yields: $\alpha =13^{\rm h}  57^{\rm m} 02\fs60$ and  $\delta  = -64^\circ 29\arcmin 29\farcs80$ with a radial error of $\sim$0\farcs38.   This implies a difference of $1\farcs23 \pm 0\farcs4$ between the \chan\ and radio-interferometry position.  
Then, we verified the  \chan\  absolute astrometry by matching the positions of serendipitous X-ray sources detected in the \chan/ACIS image with those of their putative counterparts detected in the 2MASS catalogue (Skrutskie et al.\ 2006). Since the accuracy of the \chan\ astrometry rapidly degrades at large off-axis angles, we restricted our search to within 4\arcmin\ from the centre of the ACIS field--of--view.  In particular, we found 2MASS matches for 4 X-ray sources in the PSR\, J1357$-$6429  field.  Although the uncertainty on the computed bore-sight correction does not yield any significant improvement on the absolute \chan\ astrometry, the source match does not show evidence of possible offsets between the optical and X-ray reference frames (both tied to the ICRF).   Thus, we conclude that the \chan\ coordinates are not affected by systematics.  Similarly, we verified the published radio-interferometry position of PSR\, J1357$-$6429 against the radio-timing position obtained through more recent observations performed by the {\em Fermi} Pulsar Timing Consortium (Smith et al.\  2008) and we found that they are consistent, although the latter is probably affected by the pulsar timing noise and has a much larger uncertainty.

In principle, the observed difference between the \chan\  and the radio-interferometry positions of PSR\, J1357$-$6429  can be, at least partially, due to its yet unknown proper motion. The time span between the epochs of the \chan\  and the ATCA observations is $\sim 7.17$  yrs.  This would  imply a proper motion  $\mu = 0\farcs17 \pm 0\farcs055$ yr$^{-1}$ along a position angle $\theta=70^{\circ} \pm 15^{\circ}$.  
We note that the  \chan\ observations of PSR\, J1357$-$6429
are distributed over a time span of $\sim 3.9$ yrs. Then, for the  assumed single-epoch \chan\ radial position uncertainty of $0\farcs6$, they would be sensitive only to a proper motion of $\ga 0\farcs55$  yr$^{-1}$ ($3 \sigma$), i.e. larger than inferred from the comparison between the average \chan\ position and the radio one. At the DM distance of $\sim 2.4$ kpc, the inferred proper motion would imply a transverse velocities of 2100$\pm$700 km s$^{-1}$. This value is high but, after accounting for the associated uncertainty,  it is not unheard of for a neutron star, as shown by the Guitar Nebula pulsar PSR\, B2224+65, whose transverse velocity could be as high as  1600 km s$^{-1}$ (e.g., Chatterjee \& Cordes 2004).  Thus, the computed   difference between the \chan\ and radio positions of PSR\, J1357$-$6429 might, indeed, result in its first proper motion measurement. 
A similar case is  that of the pulsar PSR\, J0108$-$1431, for which the comparison between its \chan\  and ATCA radio-interferometry positions also yielded a first evidence of a $3 \sigma$ proper motion (Pavlov et al.\ 2009), soon after confirmed by VLBI observations  (Deller et al.\ 2009). 
As in that case, future radio-interferometry observations will be crucial to confirm our tentative proper motion measurement of PSR\, J1357$-$6429.

 \begin{table}
\begin{center}
  \caption{Summary of the PSR\, J1357$-$6429 and PSR\, J1048$-$5832 coordinates (top and lower half, respectively) available from the literature from \chan\ and radio timing (T) or interferometric (I) observations and the associated radial errors  $\delta r$ at the reference epoch (fourth column). }
\begin{tabular}{llllr} \\ \hline
$\alpha_{J2000} ^{(hms)}$ &  $\delta_{J2000}^{(\circ ~'~")}$  & $\delta~r ^{(")}$ & MJD &  Source \\ \\\hline
13 57 02.42   & -64 29 30.20  & 0.15 & 51785  &  Radio,I   (1) \\	
13 57 02.54   & -64 29 30.00  & 0.60 & 53693  &  \chan\  (2) \\	
13 57 02.60   & -64 29 29.80  & 0.38 & 54402  &  \chan\  (3) \\     \hline
10 48 12.20   & -58 32 05.80 & 1.21 & 50889  &  Radio,T (4) \\
10 48 12.604 & -58 32 03.75 & 0.08 & 50581  &  Radio,I (5) \\
10 48 12.64  & -58 32  03.60 & 0.55 & 52859  &  \chan\ (6)  \\ \hline                 
\label{coo}
\end{tabular}
\end{center}
(1) Camilo et al.\ (2004); 
(2) Zavlin\ (2007);
(3) This work;
(4) Wang et al.\ (2000);
(5) Stappers et al.\ (1999);
(6) Gonzalez et al.\ (2006)
\end{table}

For PSR\, J1048$-$5832,  the \chan\ position is also quite different ($4\farcs1 \pm 1\farcs3$) from the radio-timing  one (Wang et al.\ 2000), obtained from  observations performed with the Australia Telescope National Facility (ATNF) Parkes radio telescope. On the other hand, the difference with respect to the radio-interferometry position obtained from observations performed with the ATCA (Stappers et al.\  1999) is 
not significant.
As done above, we verified all the available sets of coordinates. Firstly, we re-computed the X-ray position from the analysis of the  \chan/ACIS observation and we found a value perfectly consistent with that published in Gonzalez et al.\ (2006). Then, we verified the  \chan\  absolute astrometry by matching the positions of 13 serendipitous X-ray sources detected in the  \chan/ACIS image with those of their putative 2MASS counterparts and, also in this case, we found no evidence of possible offsets between the optical and X-ray reference frames, implying that, also in this case, the \chan\ coordinates are free from systematics. As done above, we compared the Wang et al.\ (2000) radio-timing position with that obtained by the {\em Fermi} Pulsar Timing Consortium and we found no obvious difference. The time span between the epochs of the \chan\  and the radio-timing observations of  PSR\, J1048$-$5832 is $\sim 5.39$ yrs, which would imply a proper motion of $0\farcs76 \pm 0\farcs24$ yr$^{-1}$. At the DM distance of $\sim 2.7$ kpc, this corresponds to a transverse velocity of 10250$\pm$3250 km s$^{-1}$, which is much larger than the most extreme values derived from the known radio pulsar velocity distribution (e.g., Hobbs et al.\ 2005). Thus,  the apparent inconsistency between the \chan\ and radio-timing positions cannot be explained by the pulsar proper motion, but by a genuine difference in the astrometry of either of the two observations.  In this respect, we note that the radio-timing position of Wang et al.\ (2000) is also inconsistent with the radio-interferometry one of Stappers et al.\ (1999), which suggests that the former might be affected by systematics and has to be taken with caution. Indeed, such an inconsistency was also noted by Wang et al.\ (2000) who attributed it to the effect of timing noise following period irregularities after a glitch.  
We note that if we assume the radio-interferometry position of PSR\, J1048$-$5832 as a first-epoch reference, the more recent \chan\ position would imply a  proper motion $\mu \la 0\farcs138$ yr$^{-1}$ ($1\sigma$) along an unconstrained position angle.

Although the origin of the inconsistency between the \chan\  and radio positions of the two pulsars is still debatable, and addressing it fully is beyond the goals of this work, we tend to favour the former and more recent ones.  
Thus, 
in the following we assume them as a reference. However,  not to overlook any potential candidate counterpart, we also prudently evaluate any object detected  in the \vlt\ images at, or close to, the radio positions of the two pulsars.

\begin{figure*}
\includegraphics[height=9cm]{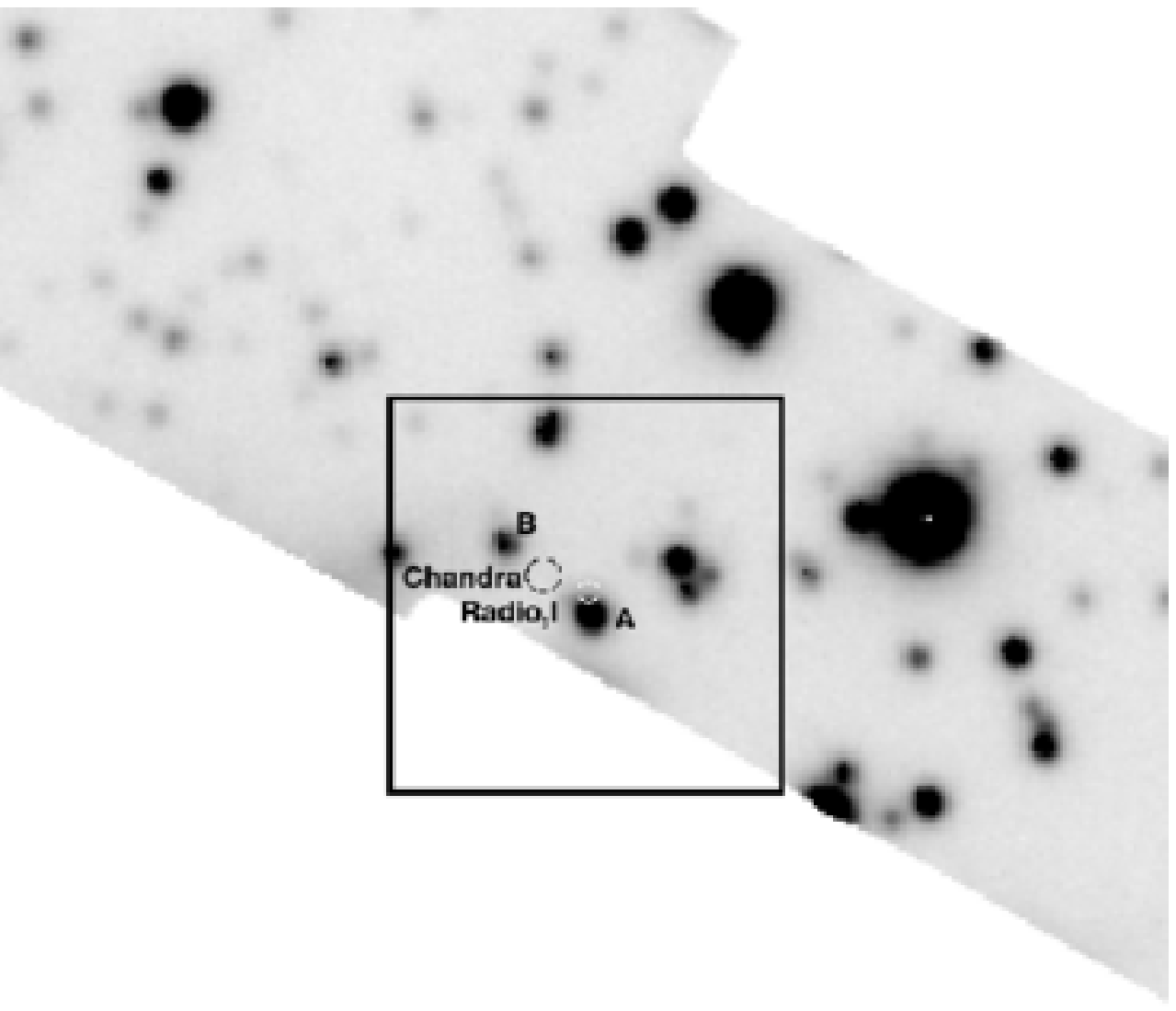}
\includegraphics[height=9cm]{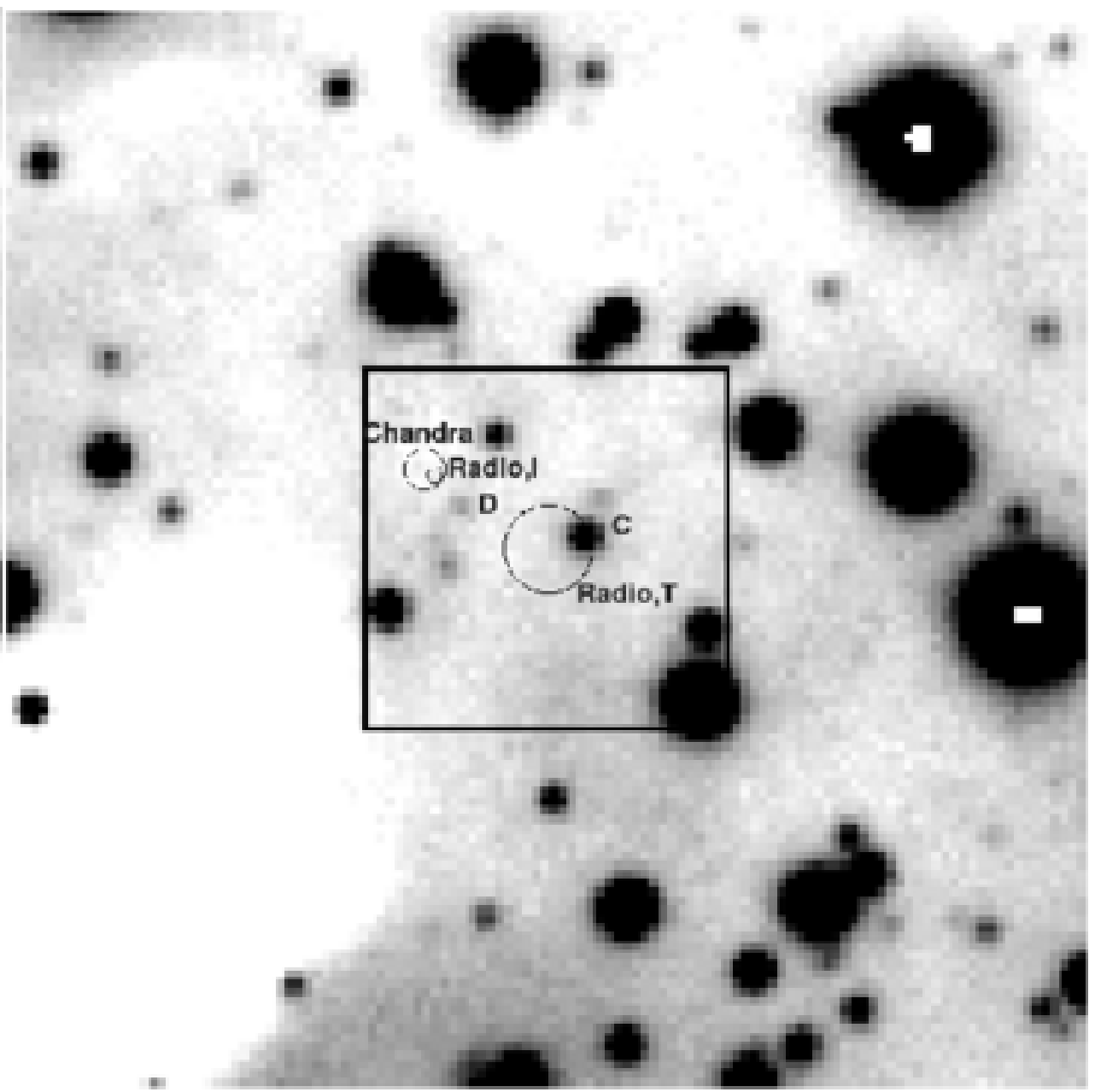}
  \caption{\vlt/\fors\ observations of PSR\, J1357$-$6429 (left)  and PSR\, J1048$-$5832 (right)  taken in the $v_{\rm HIGH}$ filter, with integration times of 14700 and 24000 s, respectively.  North to the top, East to the left.   Image cutouts are $30\arcsec \times 30\arcsec$ in size. The circles marks the pulsar radio (T=timing; I=interferometric) and \chan\ positions derived according to our astrometry re-calibration (Sect 2.2). Their radius account  for the absolute error on the reference X-ray/radio coordinates (Table\ 2) and  the accuracy of our optical astrometry (0\farcs2).
The radius of the error circles is $\sim$0\farcs43 and $\sim$0\farcs25 for the \chan\ and radio-interferometry coordinates of PSR\, J1357$-$6429, respectively, and $\sim$0\farcs58, $\sim$0\farcs22, and $\sim$1\farcs22, for the \chan, radio-interferometry, and radio-timing coordinates of PSR\, J1048$-$5832, respectively.   The white areas in the left panel correspond to the edges of the occulting bars used in the \fors\ frame to mask bright stars close to the pulsar position.  The clumpy white structure in  the right panel is part of an extended nebulosity detected in the PSR\, J1048$-$5832  field (see Sect. 3.2). In both panels, the square ($10\arcsec \times 10\arcsec$) corresponds to the area shown in Fig. 3.}
\end{figure*}

\section{Data analysis and results}

The computed positions of PSR\, J1357$-$6429 and PSR\, J1048$-$5832 are shown in Fig. 1, overlaid on the stacked $V$-band \fors\ images of the two pulsars. The overall uncertainties on the pulsar positions account for the absolute error on the reference \chan\ and radio coordinates (Table 2) and the accuracy of our astrometry re-calibration of the \fors\ images (0\farcs2). However, to conservatively account for the unknown pulsar proper motions, in the following we extended our search radius up to 3 times  the estimated $1 \sigma$ uncertainties. 

\subsection{PSR\, J1357$-$6429}

The radio position of  PSR\, J1357$-$6429 (Fig. 1, left panel) falls at $\sim 0\farcs65$ from  that of a relatively bright field star (Star A; $V=21.8 \pm 0.06$), i.e at an angular distance equal to the $\sim$ 2--3 $\sigma$ radio position uncertainty.
To quantitatively assess the significance of the association, we estimated the  chance coincidence probability that an unrelated field object falls within a radius of 0\farcs65 from the computed pulsar radio position.  This probability can  be  computed as  $P=1-\exp(-\pi\sigma r^2)$,  where  $\mu$ is  the measured object density in the \fors\  upper CCD  field--of--view (accounting for areas affected by vignetting and the areas masked by the occultation bars, see Sect. 2.1) and $r$ is  the matching radius  ($0\farcs65$).  The density of star-like  objects (ellipticity  $e<0.2$) with  magnitude  $V\ga 22$  in  the field--of--view  is  $\mu \sim$ 0.005/sq. arcsec. This yields an estimated chance coincidence  probability $P \sim  9 \times 10^{-3}$, not statistically compelling yet. 

In order to qualitatively verify the reliability of such an association, we compared the brightness of Star A with that expected from the pulsar.  PSR\, J1357$-$6429 has a rotational energy loss $\dot{E} \sim 3.1 \times 10^{36}$ erg s$^{-1}$. If we assume an efficiency in converting spin-down power into optical luminosity $\eta_{opt} \equiv L_{opt}/\dot{E}$ comparable to that of, e.g. the $\sim 11$ kyr old Vela pulsar, we would expect an optical luminosity for PSR\, J1357$-$6429 lower by only a factor of $\sim 2$, after scaling for the Vela rotational energy loss ($\dot{E} \sim 6.9 \times 10^{36}$ erg s$^{-1}$). At the PSR\, J1357$-$6429 distance  ($2.4 \pm 0.6$ kpc) and  for the corresponding interstellar extinction ($A_V = 2.2^{+1.7}_{-1.1}$), computed from the 
$N_H=0.4^{+0.3}_{-0.2} \times 10^{22}$ cm$^{-2}$ obtained from the spectral fits to the \xmm\ spectrum (Esposito et al.\ 2007) using the relation of Predhel \& Schmitt\ (1995) and the extinction coefficients of Fitzpatrick\ (1999), we then derive an expected magnitude in the  range $V\sim 29.5$--33.4, accounting for both the distance and interstellar extinction uncertainties. Obviously, these values are not within reach for any current 10m-class telescope. 
On the other hand, if  PSR\, J1357$-$6429 had an efficiency comparable to that of the younger  ($\tau \sim 1$ kyr)  and more energetic ($\dot{E} \sim 4.6 \times 10^{38}$ erg s$^{-1}$) Crab pulsar, the same scaling as above would yield an expected magnitude of $V\sim 21.1$--24.9.
Thus, if Star A were the pulsar's optical counterpart, PSR\, J1357$-$6429 would have an emission efficiency either comparable or up to $\sim 15$ times larger than the Crab.  Thus, the association would not be impossible, in principle, at least based on the pulsar's energetics and an assumed Crab-like optical emission efficiency. 
However, we note that pulsar optical emission efficiencies have been computed for less than 10 objects and their dependance on the pulsar parameters, such as their spin-down age, is yet unclear. Therefore, we  followed an independent approach to estimate the expected pulsar's brightness using as a reference the relation between the non-thermal X-ray and optical luminosities measured for pulsars with identified X-ray/optical counterparts (e.g., Zharikov et al.\  2004; 2006).  In this case, taking into account the rather large uncertainty affecting this relation, as well as the uncertainty on the X-ray spectral parameters of  PSR\, J1357$-$6429 (Esposito et al.\ 2007), hence on the non-thermal X-ray luminosity, we end up with an expected magnitude in the  range $V\sim$26.2--31.2, i.e. much fainter  than that of star A.
As a further test, we also compared the colours of Star A ($V=21.80 \pm 0.06$; $R=20.65 \pm 0.03$; $I=19.75\pm 0.03$) with those of field stars, as measured  on the short exposures to avoid problems in background subtraction for stars close to the occulting bars, saturation problems, and to homogeneously cover the unmasked \fors\ field--of--view. The two $V$-$(V-R)$ and $V$-$(V-I)$ colour-magnitude diagrams (CMDs) are shown in Fig. 2. As seen, the  location of Star A in both CMDs is along the sequence of field stars, implying that it has no peculiar colours and that its association with the pulsar is unlikely.  For instance, for a flat power-law spectrum $F_{\nu} \propto \nu^{-\alpha}$, like that of, e.g. the Vela pulsar (Mignani et al.\ 2007), we would expect a $(V-R) \approx 0.2$ and a $(V-I) \approx 0.1$, once accounting for the interstellar extinction towards PSR\, J1357$-$6429, which are obviously off the main sequence in both CMDs.
Thus, on the basis of the comparison with the optical emission efficiency of pulsars with known optical counterparts, their relative X-ray--to--optical brightness, and the colours of Star A, we consider it unlikely that it is the pulsar's counterpart.

\begin{figure*}
\includegraphics[height=9cm]{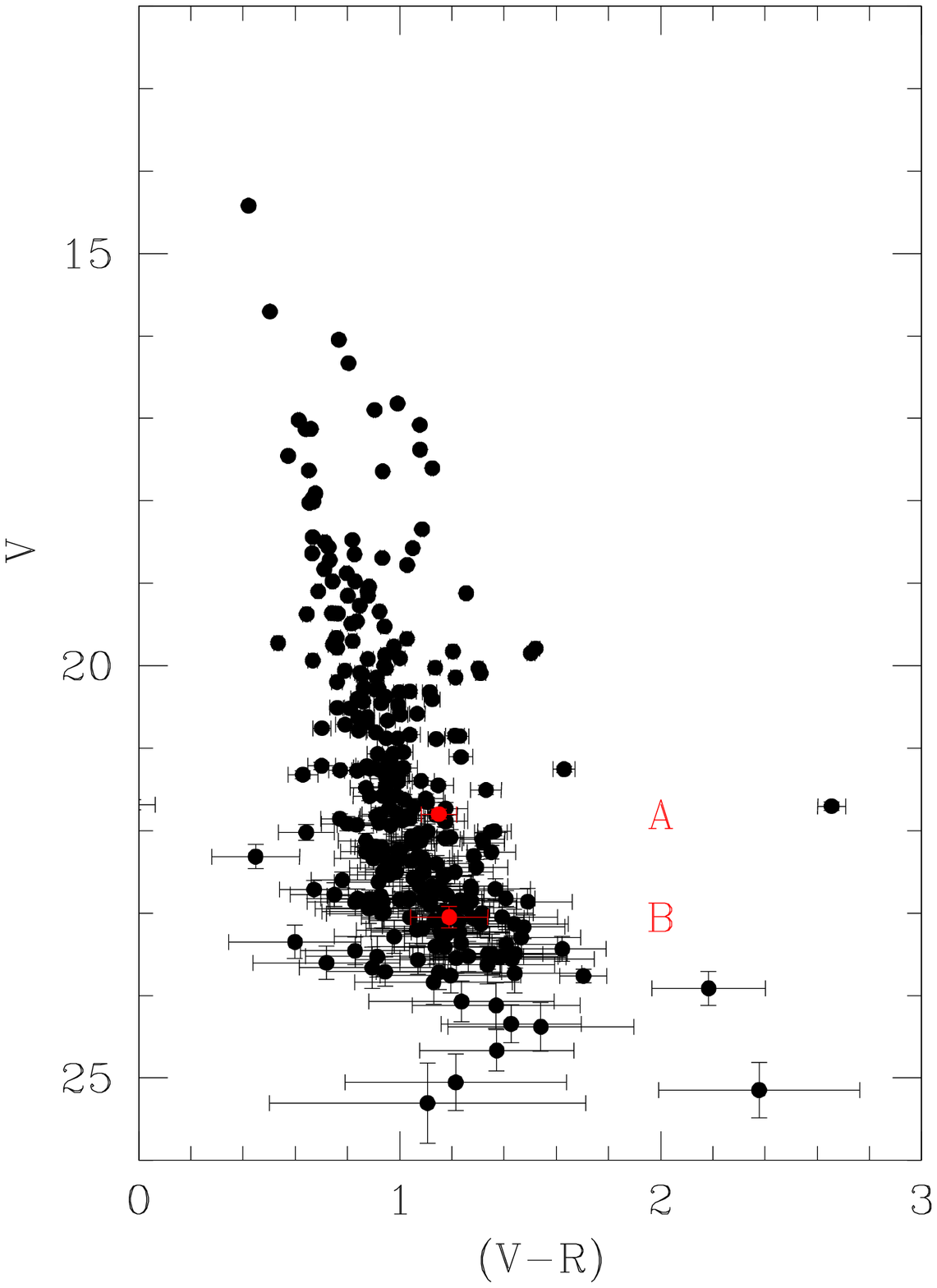}
\includegraphics[height=9cm]{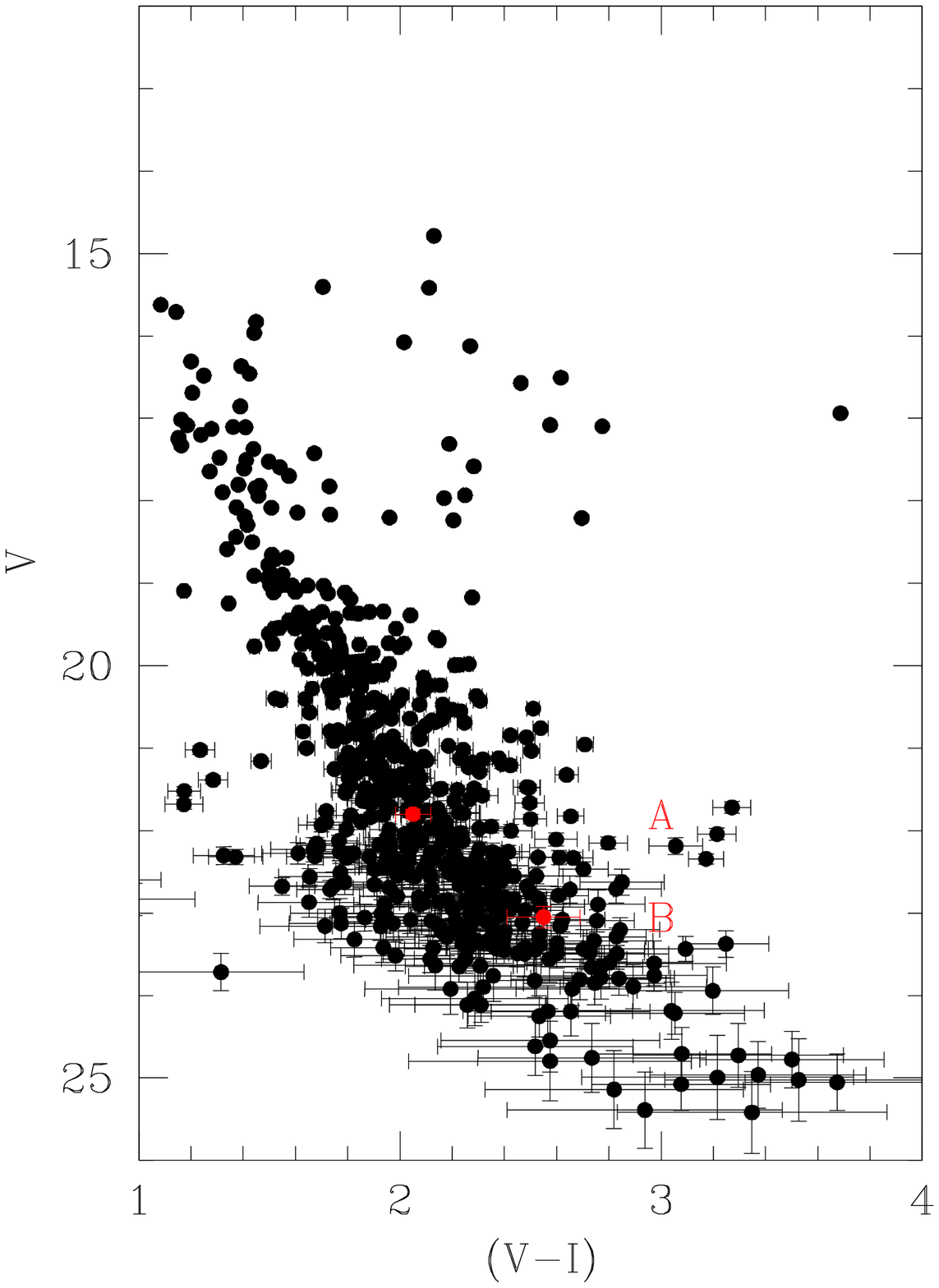}
  \caption{Observed (not extinction-corrected) colour-magnitude diagrams of all the objects detected in the PSR\, J1357$-$6429 field. The two objects detected closest to the radio (Star A) and \chan\ (Star B) pulsar positions (Fig. 1) are marked in red, with magnitudes  $V=21.8 \pm 0.06$ and $V=23.05 \pm 0.13$, respectively. The photometry errors are purely statistical and do not account for the systematic uncertainties on the absolute flux calibrations.}
\end{figure*}

We then searched for possible counterparts at our revised \chan\ position. We note that this overlaps with an apparent enhancement over the sky background noticed in the co-added $I$-band image (Fig. 3, left).
However, it is difficult to determine whether such an enhancement is due to a background fluctuation, perhaps produced by the superposition of the PSF wings of the two adjacent stars detected southwest (Star A) and northeast (Star B) of the \chan\ position  and of the bright masked star southeast of them, or it is associated with a real source. 
Unfortunately, the difficulties in the PSF subtraction of the two stars, with Star A possible blended with a fainter star and Star B at the edge of occulting bar and partially blended with a bright star, do not allow us to better resolve the background enhancement.  If associated with a point source, such an enhancement would correspond to a magnitude $I\approx 24.6$. Interestingly enough, such an enhancement is recognisable only in the $I$-band image and not in the longer-integration $R$ and $V$-band ones. This would suggest that either it is not associated with a real object or, if it is,  that the object's spectrum is either quite red and/or affected by an interstellar extinction probably larger than measured in the pulsar's direction.  To verify the presence of an object at the \chan\ position, we inspected the single $I$-band images. However,  their short integration time (200 s) makes it difficult to recognise any obvious systematic flux enhancement at the expected location.
We also smoothed the images using a Gaussian filter  over $3\times3$ pixel cells, but this did not yield to a clearer detection. 
Thus,  it is difficult to prove that the background enhancement seen at the pulsar \chan\ position in the $I$-band image is unambiguously associated with a real object.
Higher spatial resolution observations with the \hst\ would be crucial to resolve this putative object against the noisy background produced by the PSF of Stars A and B and confirm it as a candidate optical counterpart to PSR\, J1357$-$6429. 

No other possible counterpart is detected within the computed pulsar \chan\ error circle (Fig.3, left). Star B  is at $\sim1\farcs4$, i.e. $\approx 3 \sigma$ from the best-estimate \chan\ position.  Moreover, both its fluxes ($V=23.05 \pm 0.13$;  $R=21.86 \pm 0.07$; $I=20.5 \pm 0.05$)  and colours (Fig. 2) are comparable with those of Star A, already ruled out as a candidate counterpart (see above). Thus, assuming that PSR\, J1357$-$6429 is not detected in the \fors\ images we computed the pulsar flux upper limits in the $VRI$ bands. Following a standard approach (e.g., Newberry\ 1991), we determined the number of counts corresponding to a $3 \sigma$ detection limit  in a photometry  aperture of 1\arcsec\ diameter (8 pixels) from the standard deviation of the background  sampled within the \chan\  error circle.   After applying the aperture correction, computed  from the measured PSF of a number of relatively bright but unsaturated stars in  the field, we then derived $3 \sigma$ upper limits of $V \sim 27$ and $R \sim 26.8$
at the \chan\ position, while in the $I$ band we conservatively assumed a flux of $I\approx 24.6$ measured above as our upper limit estimate.

\begin{figure*}
\includegraphics[height=9cm]{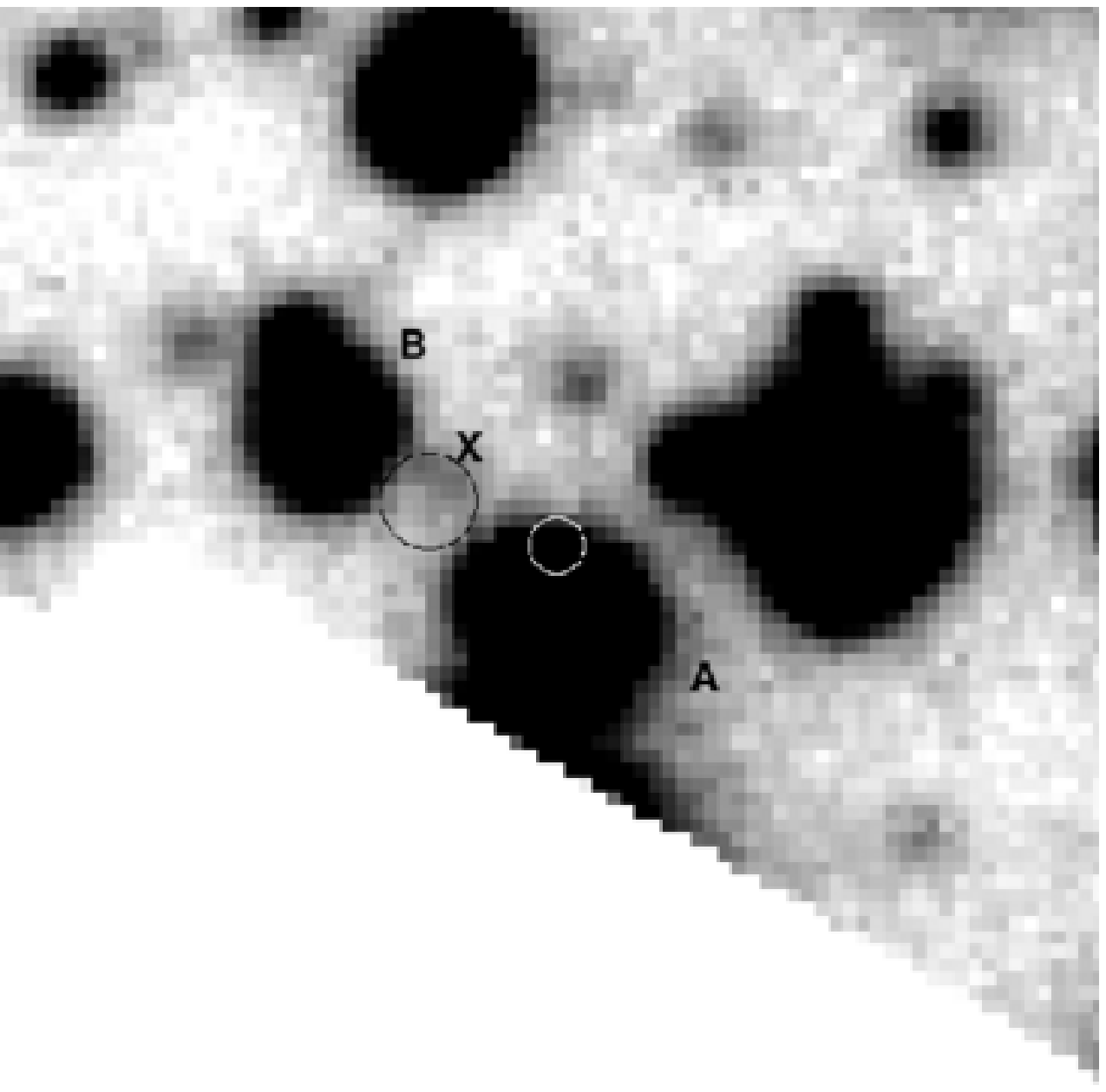}
\includegraphics[height=9cm]{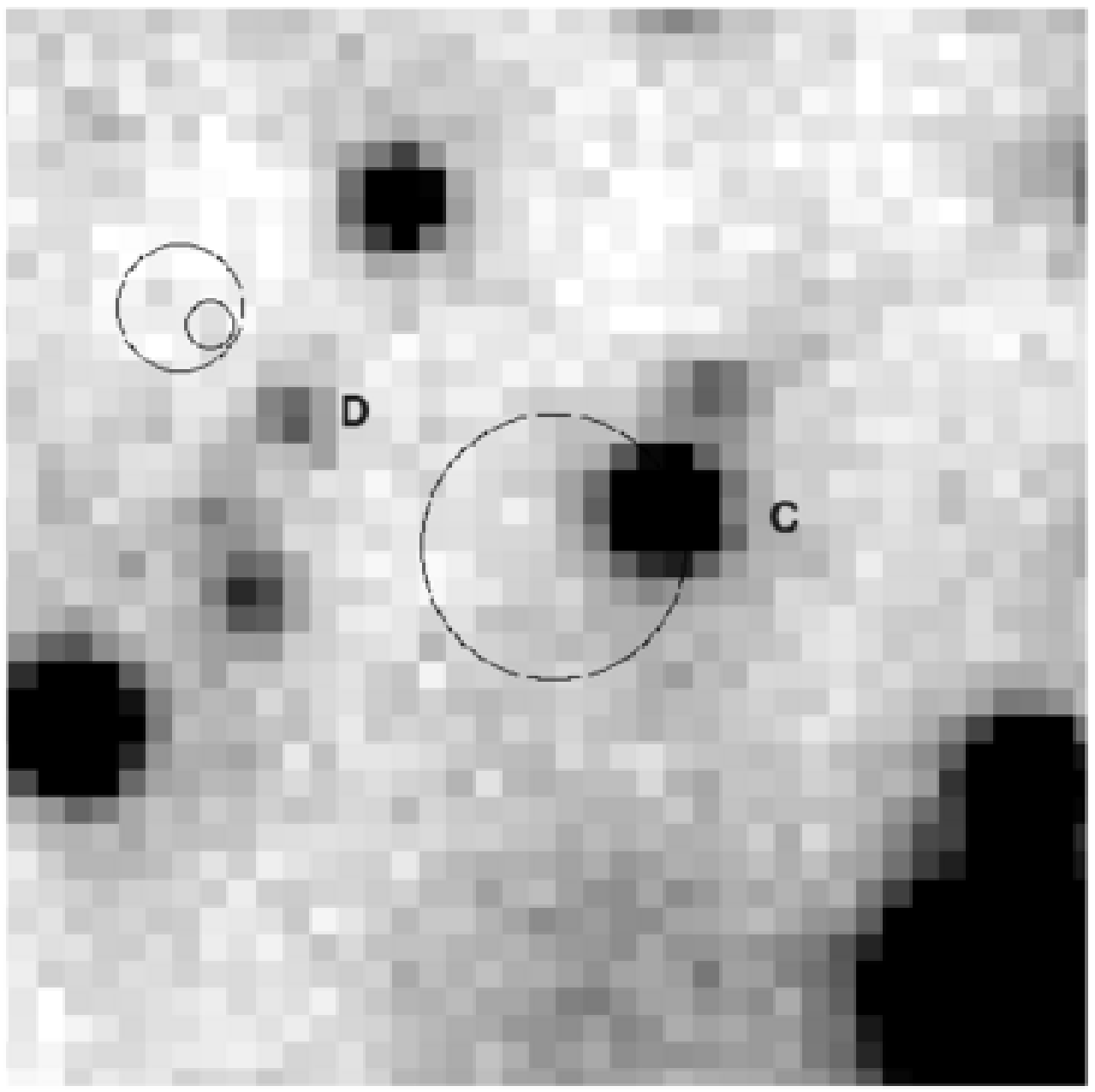}
  \caption{$10\arcsec \times10\arcsec$ zooms of the PSR\, J1357$-$6429 (left) and  PSR\, J1048$-$5832 images (right) taken through the  $I_{\rm BESS}$ (1800 s) and the  $v_{\rm HIGH}$ filter (24000 s), respectively. In both cases,  the colour scale has been stretched to highlight fainter objects. A flux enhancement (labelled with X) is seen at the \chan\ position of PSR\, J1357$-$6429, which might be due either to a background fluctuation or to an unresolved object in the PSF of the two adjacent stars A and B (see text for discussion). }
\end{figure*}

\subsection{PSR\, J1048$-$5832}

The field of PSR\, J1048$-$5832 shows large extended structures where both the sky brightness and the star density are much lower then in the rest of the field.  These structures are present in the single \fors\ raw frames taken days apart so that they are not an artefact due to, e.g. an incorrect flat fielding or to the unreported presence of clouds at the moment of the observations. We have checked the $R$-band images of the Digitised Sky Survey (DSS) and we found that the same structures are visible with the same extent and position as seen in the \fors\ images, which proves that they are real and not due to any instrumental or atmospheric effects. We used the DSS to investigate the presence of such structures on scales as large as  $40\arcmin \times 40\arcmin$ and we found that they apparently connected to other similar structures extending on a sort of regular pattern.  We have also checked the 2MASS $JHK$-band images which, instead, do not show either evidence of such structures or of under-density of stars at their expected locations. Thus, we conclude that they are most likely dense molecular clouds along the plane of the Milky Way, possibly part of the CH$_{\rm 3}$OH maser associations in nearby star-forming regions (e.g. Goedhart et al.\ 2004).

Although no counterpart to PSR\, J1048$-$5832 has been yet proposed in the literature, we note that an observing program\footnote{{\it Confirming the detection of a Fermi gamma-ray pulsar in the optical},  PI Sollerman, 386.D-0585(A).} has been recently carried out at the \vlt\ to follow-up on the claimed detection of a candidate counterpart, presumably in the same data set used in this work. 
This prompted us to prudently screen each object detected close to the published positions of PSR\, J1048$-$5832.  In particular, we note that its radio-timing position (Fig. 1, right panel) falls close to a $V\sim 24$ object (Star C), even detected in the short 15s exposures used for the image astrometry,  right on the edge of the radio error circle ($\sim 1\farcs22$).   Again, based on position alone we can not rule out that Star C is associated with the pulsar.  However,  the probability of chance coincidence with field objects of magnitude $V\ga 24$  is  $P\sim 0.04$, i.e. certainly not low enough to statistically claim an association.  In addition, such an association is problematic from the point of view of the pulsar's energetics. As done in Sect. 3.1,  we compared the brightness of Star C with that expected for the pulsar.  
By assuming an optical efficiency  comparable to that of the Vela pulsar, PSR\, J1048$-$5832's rotational energy loss ($\dot{E} \sim 2 \times 10^{36}$ erg s$^{-1}$) would then imply  only a factor of $\sim 3$ lower optical luminosity.  Following the same analysis as for PSR\, J1357$-$6429 (see Sectn. 3.1),  at the PSR\, J1048$-$5832 distance  ($2.7\pm 0.35$ kpc) and for the corresponding interstellar extinction ($A_V = 5^{+2.2}_{-1.1}$), derived from the  $N_H=0.9^{+0.4}_{-0.2} \times 10^{22}$ cm$^{-2}$  (Marelli et al.\ 2011), we determine an expected magnitude $V\sim33.3$--37,  accounting for both the distance and  interstellar extinction uncertainties.
Similarly, assuming more optimistically an emission efficiency comparable to that of the Crab pulsar, we derive an expected magnitude of $V\sim 25$--28.7.
This means that, if Star C  were indeed the PSR\, J1048$-$5832 optical counterpart, the pulsar should have an emission efficiency $\sim 2$--80 times larger than the Crab. On the other hand, by using the relation between the X-ray and optical luminosity (Zharikov et al.\ 2004; 2006) as done in Sectn. 3.1, we obtain a magnitude in the  range $V\sim$ 29--34.  Thus,  Star C cannot be the pulsar optical counterpart. 

Then, we  searched for a possible counterpart at the pulsar \chan\  position. No object is detected within, or close to, the  \chan\ (0\farcs55) error circle apart from a faint object (Star D; $V\sim 26.7$) visible $\sim 1\farcs6$ southwest of it (Fig.3, right). However, the offset is about  3 times the $1	 \sigma$  uncertainty on the pulsar position. Thus, we deem the association unlikely both on the basis of the loose positional coincidence and on statistical grounds, with a chance coincidence probability $P\sim 0.08$.   Moreover, the lack of colour information makes it impossible to constrain the nature of this object. Therefore, we conclude that PSR\, J1048$-$5832  is not detected in the \fors\ image.
Following the same procedure as used in Sect. 3.1, we determined the number of counts corresponding to a $3 \sigma$ detection limit in a 1\arcsec\ photometry aperture (4 pixel) from the standard deviation of the background sampled within the \chan\ error circle. After applying the aperture correction, this yield to a $3\sigma$ detection limit of $V\sim 27.6$.

\begin{figure*}
\includegraphics[height=7cm]{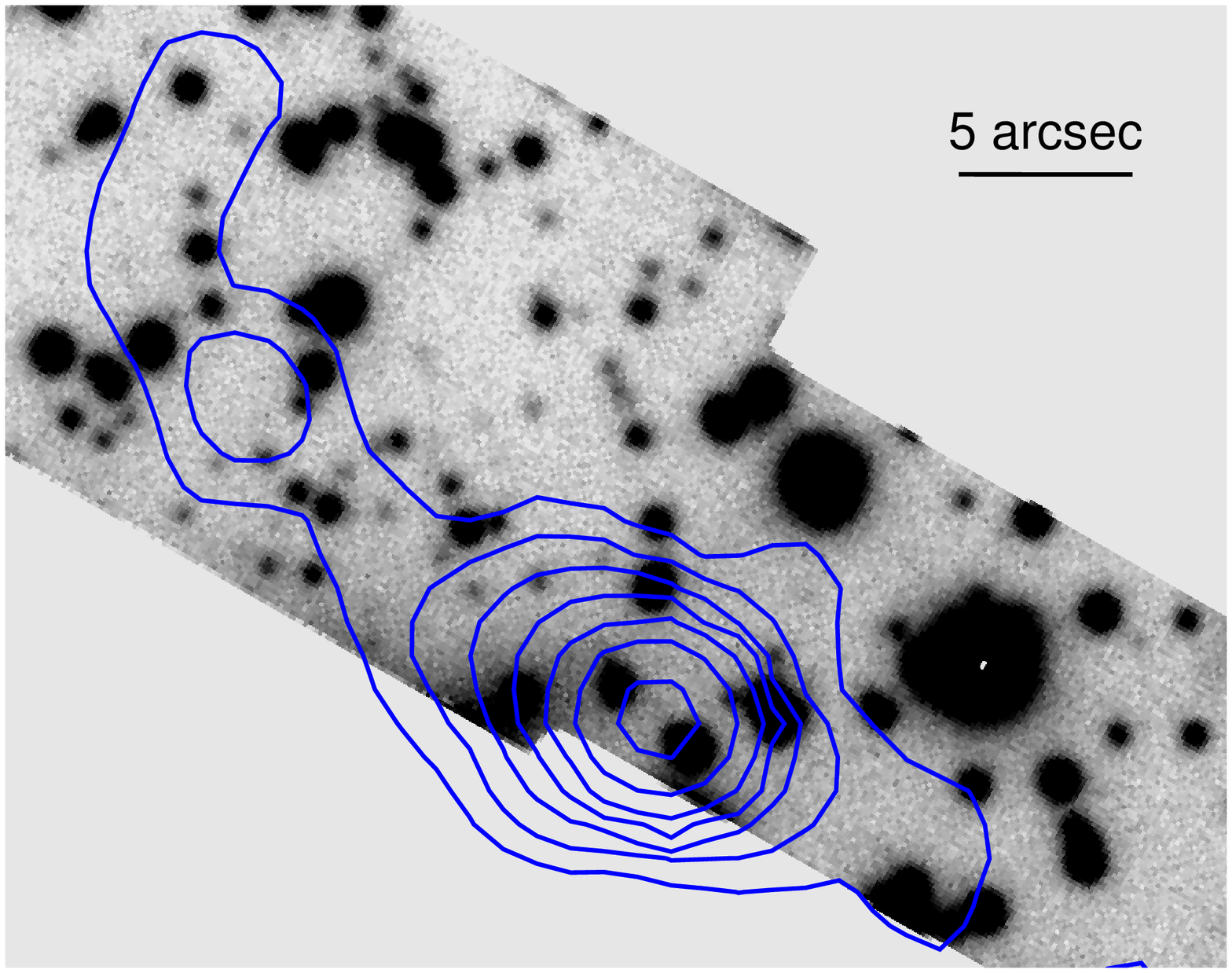}
\includegraphics[height=7cm]{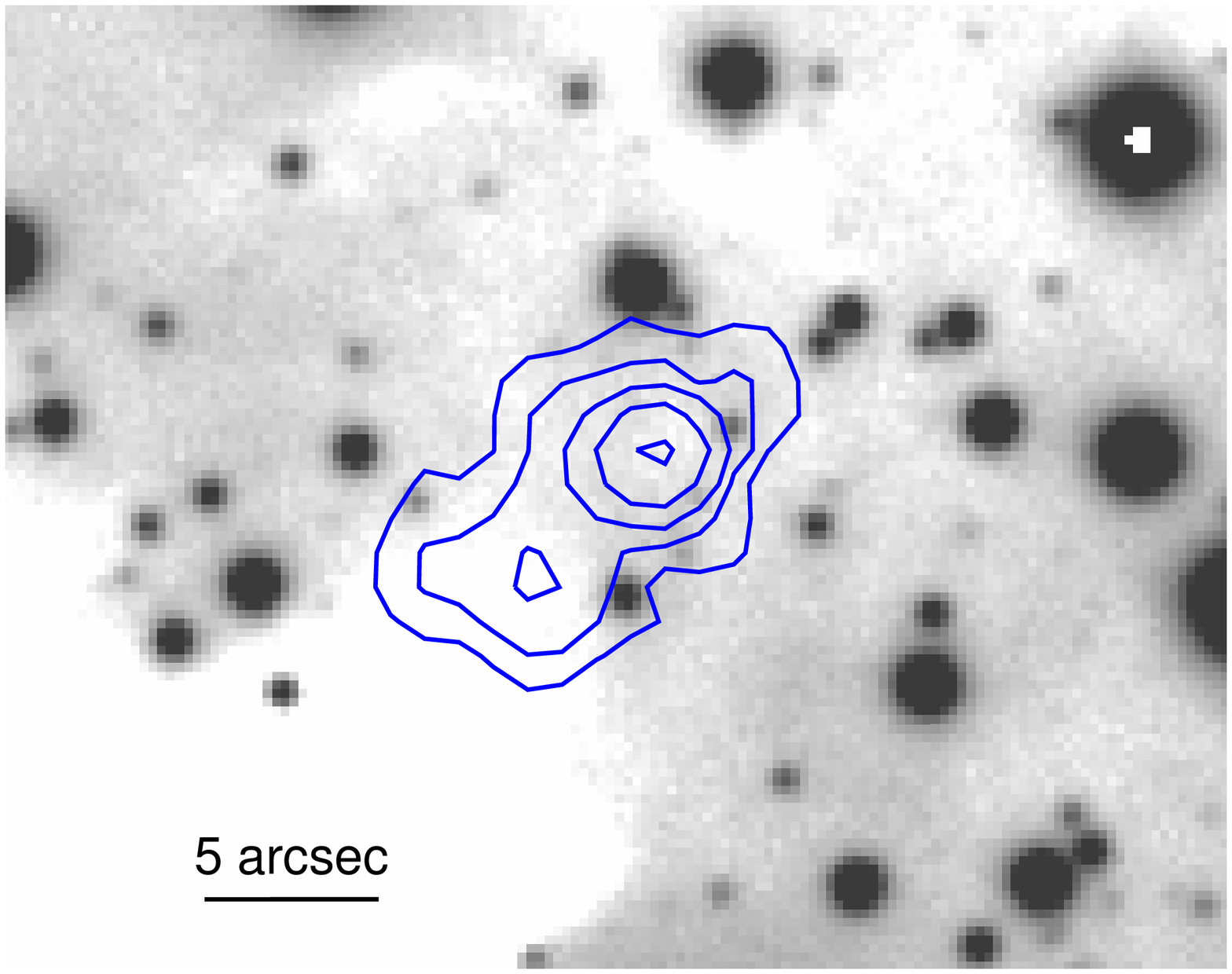}
  \caption{Zooms of the PSR\, J1357$-$6429 (left) and  PSR\, J1048$-$5832 (right)  \fors\  $V$-band images. The blue contours corresponds to the isophotes underlying the structure of the PWNe observed by \chan\ in the 0.5--4 keV range. In both cases, a  smoothing with  a Gaussian kernel of $1\arcsec$ has been applied.  The contours have logarithmic spacing, with a factor of 2 step in surface brightness. }
\end{figure*}

\subsection{Search for the optical PWNe}

In order to verify the possible presence of PWNe in the optical, we have over plotted the X-ray contours of the \chan/ACIS images of  PSR\, J1357$-$6429 and PSR\, J1048$-$5832 on the \fors\ $V$-band images (Fig. 4).  The X-ray PWNe of PSR\, J1357$-$6429 has been discovered in a recent  $\sim$ 59 ks \chan/ACIS image (Lemoine-Goumard et al.\ 2011; Chang et al.\ 2011) and has an angular extent much larger than that of the compact PWN ($\la 4\arcsec$) tentatively detected by Zavlin\ (2007) in an older $\sim 33$ ks \chan/HRC  data set. We note that the proximity of the  PSR\, J1357$-$6429 position to the \fors\ occulting bars and the relative crowding of the field makes it very difficult to search for the optical counterpart of its  X-ray PWN.  In particular, the bright core of the PWN overlaps with the position of Stars A and B and it is partially masked by the occulting bars south of it. Due to the  relatively high fluxes of  Stars A and B with respect to that expected for a putative optical PWN,  the uncertainty in the PSF subtraction residuals makes its detection improbable.   At the same time, the faint PWN tail overlaps with several stars detected northeast of the pulsar's \chan\ position (Fig. 4, left).  Thus, any upper limit on the optical surface brightness of the PWN would be highly uncertain and hampered by the partially covered area.  We note that a PWN around PSR\, J1357$-$6429 has been detected at TeV energies by HESS (Abramowski et al.\, 2001), but its angular size is much larger than the entire \fors\ field--of--view, by itself masked by $\sim 50\%$ by the occulting bars.
In the case of PSR\, J1048$-$5832, the proximity of the \chan\ position to one of the clumps belonging to the large molecular cloud complex detected in the field (Fig.4, right) makes it difficult to search for optical emission along the whole PWN,  whose angular extent ($6\arcsec \times 11\arcsec$; Gonzalez et al.\ 2006) is partially covered by the clump. In particular, the clump entirely covers the PWN tail, which extends southeast of the pulsar position.  No extended optical emission is recognised close to the head of the PWN, where the clumps are sparser and smaller. As in the case of PSR\, J1357$-$6429, any upper limit of the optical surface brightness of the PWN is hampered by the covered area.

\begin{figure*}
\includegraphics[height=6.5cm,angle=0,clip=]{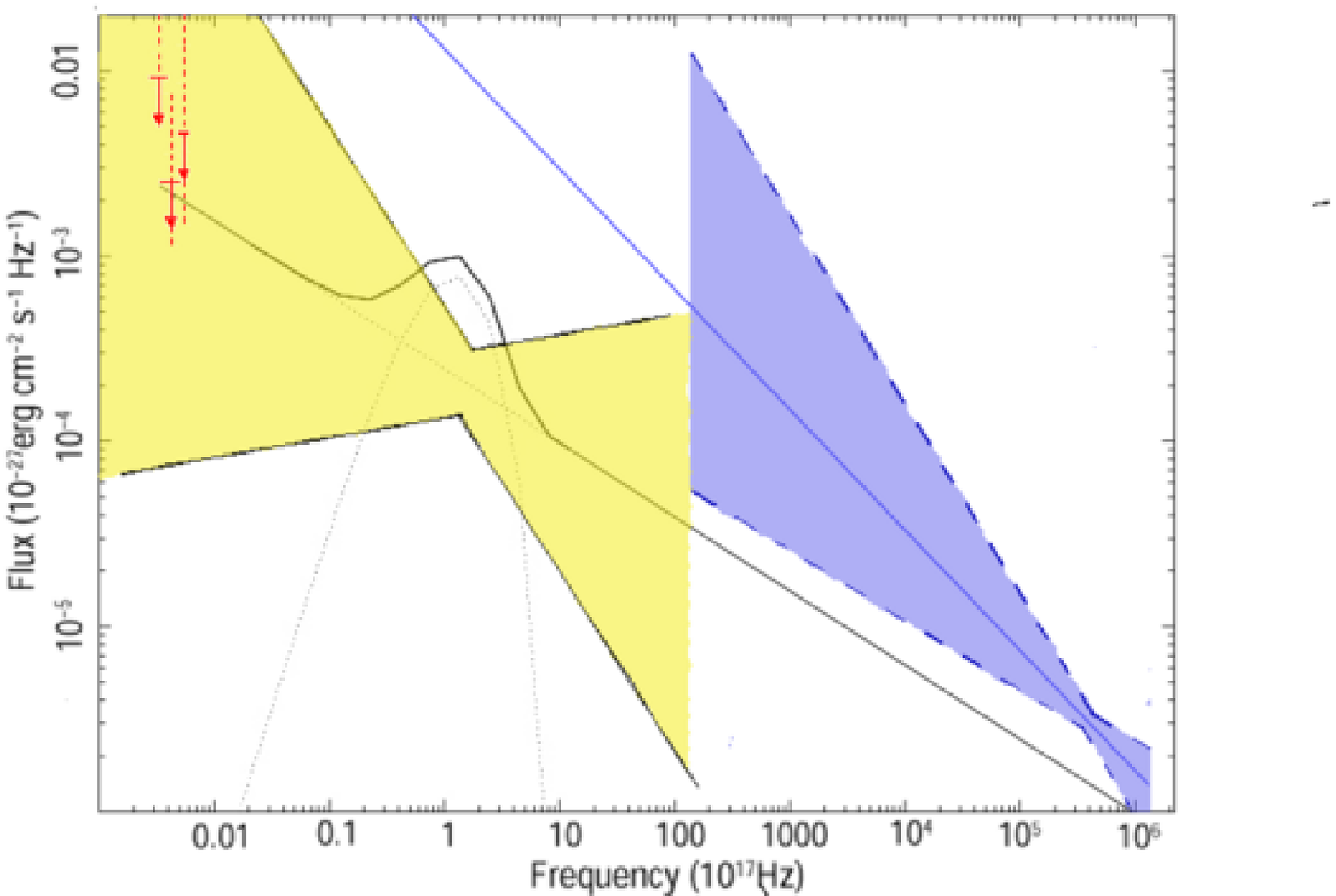}
\includegraphics[height=6.5cm,angle=0,clip=]{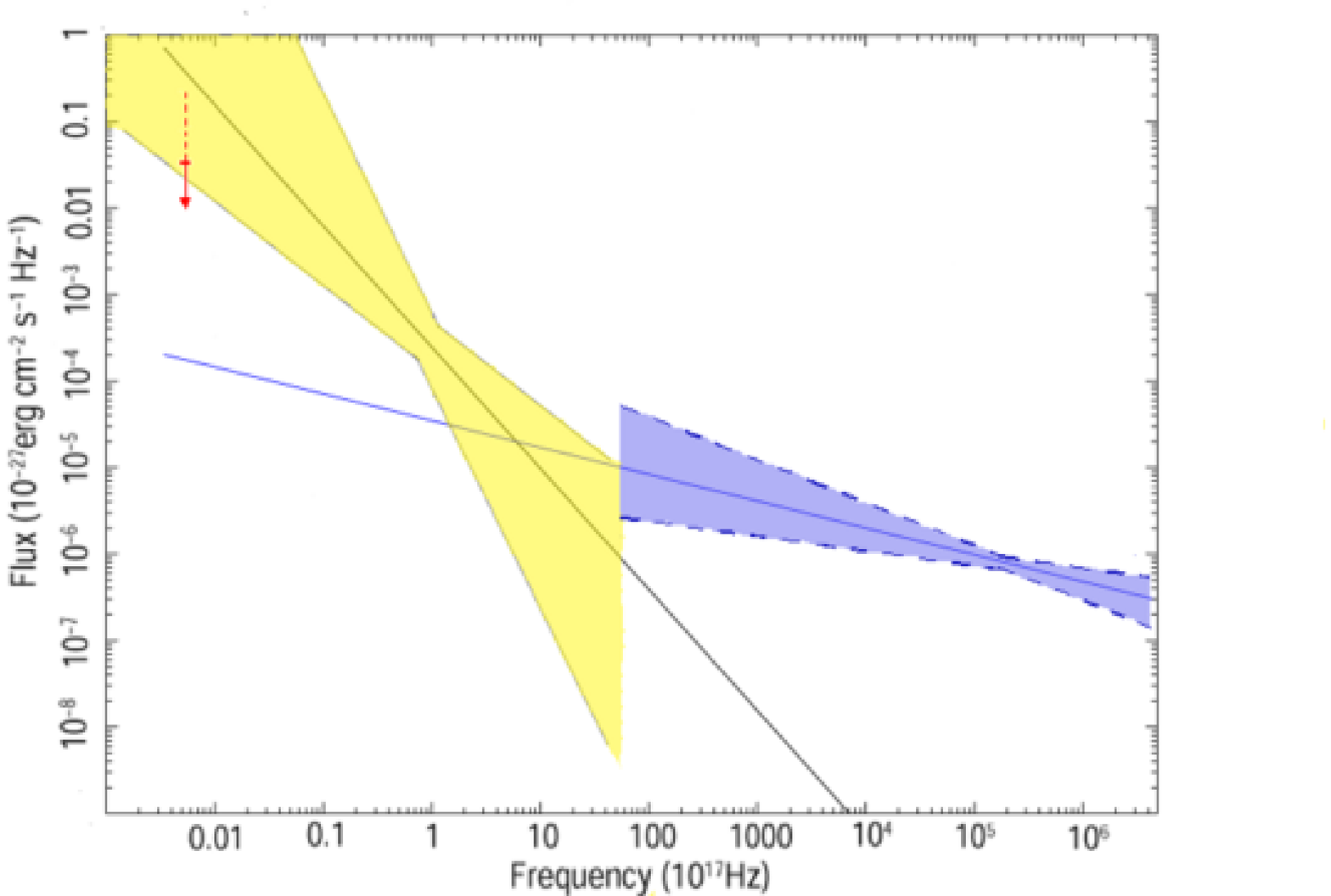}
\caption{Upper limits on the extinction-corrected optical fluxes of PSR\, J1357$-$6429 (left) and  PSR\, J1048$-$5832 (right)  compared with the low-energy extrapolations of the X-ray (solid black) and $\gamma$-ray (solid blue) spectral models which best fit the \xmm, \chan\ (Esposito et al.\ 2007; Marelli et al.\ 2011), and {\em Fermi} (Lemoine-Goumard et al.\ 2011;  Abdo et al.\ 2009) data.  The plotted optical flux upper limits are corrected for interstellar extinction based upon the best-fit value of the $N_H$. The dotted  lines in the left panel correspond to the PL and BB components to the model X-ray spectrum.  In both panels, the yellow and blue-shaded areas (left to right) indicate the $1 \sigma$ uncertainty on the extrapolations of the X and $\gamma$-ray PLs, respectively.  The vertical red dashed lines mark the uncertainty on the extinction-corrected optical flux upper limits, computed around the best-fit value of the $N_H$. For  PSR\, J1357$-$6429, we neglected the uncertainties on the extrapolation of the BB component, whose contribution in the optical band is negligible.
}
 \end{figure*}

\section{Discussion}

We compared our optical flux upper limits in the $V$ band with the pulsars' rotational energy loss rates.  For PSR\, J1357$-$6429, our upper limit of $V\sim 27$ corresponds to an optical luminosity upper limit $L_{opt} \sim 0.6$--$ 21.6 \times 10^{29}$ erg~s$^{-1}$,  for a distance $d=2.4 \pm 0.6$ kpc and for an interstellar extinction $A_V = 2.2^{+1.7}_{-1.1}$, after accounting for their associated uncertainties. This implies  an emission efficiency upper limit $\eta_{opt} \sim 0.2$--$7 \times 10^{-7}$.  This value is at least a factor of  5 lower than the Crab pulsar and, possibly, closer to that of the Vela pulsar.  
On the other hand, for PSR\, J1048$-$5832 our upper limit of $V\sim 27.6$ implies (for $d=2.7\pm 0.35$ kpc and $A_V = 5^{+2.2}_{-1.1}$) upper limits of  $L_{opt} \sim 0.4$--$12.5 \times 10^{30}$ erg~s$^{-1}$ 
 and $\eta_{opt} \sim 1.8$--$62.5 \times 10^{-7}$.  In principle, this does not rule out an optical emission efficiency comparable to that of the Crab pulsar, although the pulsar spin-down age (20.3 kyr) might suggest, also in this case,  a Vela-like emission efficiency.  
An optical emission efficiency $\eta_{opt} \la 10^{-7}$--$10^{-6}$ has been measured also from the upper limit on the optical emission of PSR\, B1706$-$44, the only other Vela-like pulsar for which \vlt\ observations are available (e.g. Mignani et al.\ 1999).  This would confirm that Vela-like pulsars are intrinsically less efficient emitters in the optical than Crab-like pulsars,   possibly even less efficient than middle-aged and old pulsars, like PSR\, B0656+14, Geminga, PSR\, B1055$-$52, PSR\, B1929+10, and PSR\, B0950+08. The measurement of such a low emission efficiency in the optical band for Vela-like pulsars might then provide useful information to emission models of the neutron star magnetosphere.

We compared the  flux upper limits of  PSR\, J1357$-$6429 and PSR\, J1048$-$5832 with the extrapolations in the optical domain of the X and $\gamma$-ray spectra.  For PSR\, J1357$-$6429, we assumed the X-ray spectral model of Esposito et al.\ (2007), a power-law (PL) with photon index $\Gamma_X=1.4\pm 0.5$ plus a blackbody (BB) with temperature $kT=0.16^{+0.09}_{-0.04}$ keV ($N_H=0.4^{+0.3}_{-0.2} \times 10^{22}$ cm$^{-2}$), and the $\gamma$-ray spectral model of Lemoine-Goumard et al.\  (2011), a PL with photon index $\Gamma_{\gamma} =1.54 \pm 0.41$ and exponential cut-off at $\sim$ 0.8 GeV. For PSR\, J1048$-$5832, we assumed the X-ray spectral model of  Marelli et al.\ (2011), a PL with photon index $\Gamma_X= 2.4 \pm 0.5$ ($N_H=0.9^{+0.4}_{-0.2} \times 10^{22}$ cm$^{-2}$), and the $\gamma$-ray spectral model of Abdo et al.\ (2009), a PL with photon index $\Gamma_{\gamma}=1.38\pm 0.13$ and exponential cut-off at $\sim$ 2.3 GeV. 
Our optical flux upper limits are corrected for interstellar extinction based upon the $N_H$ derived from the fit to the X-ray spectra.
The multi-wavelength spectral energy distributions (SEDs) of the two pulsars are shown in Fig. 5, where we accounted for both the $1 \sigma$ uncertainty on the extrapolations of the X and $\gamma$-ray PL and the uncertainty on the extinction-corrected flux upper limits. In the case of PSR\, J1357$-$6429 (Fig.5, left) we see that the optical flux upper limits can be compatible with the extrapolation of the X-ray PL. Thus,  it is possible that the expected optical PL spectrum indeed follows the extrapolation of the X-ray one, a case so far observed only for PSR\, B1509$-$58 among all the optically-identified pulsars (see, e.g. Mignani et al.\ 2010a).
The optical flux upper limits are also compatible, with the possible exception of the R-band one, with the extrapolation of the  $\gamma$-ray PL, which does not allow us to prove that there is a break in the optical/$\gamma$-ray spectrum, as observed in other pulsars.  
Indeed, a possible consistency between the $\gamma$-ray and optical PL spectra has been found so far for a minority of cases only (Mignani et al., in preparation) like, e.g. the middle-aged pulsar PSR\, B1055$-$52 (Mignani et al.\ 2010b).  To summarise, we can not rule out that a single model can describe the optical--to--$\gamma$-ray magnetospheric emission of PSR\, J1357$-$6429. The  multi-wavelength SED is different  in the case of PSR\, J1048$-$5832 (Fig. 5, right), for which the optical V-band upper limit  is compatible with the extrapolation of the  steep  X-ray PL  only for the largest values of the $N_H$ but is well above the extrapolation of the   flat $\gamma$-ray one.  This does not rule out that there is a break in the optical/X-ray PL spectrum, as observed in most pulsars (Mignani et al.\ 2010a),
and that the optical spectrum follows the extrapolation of the  $\gamma$-ray PL.
Interestingly enough, at variance with the case of PSR\, J1357$-$6429,
no single model can describe the optical--to--$\gamma$-ray magnetospheric emission of PSR\, J1048$-$5832.

Thus, the comparison between the  multi-wavelength  SEDs of these two Vela-like pulsars 
suggests that the occurrence of spectral breaks might not correlate with the pulsar's age. Of course, the uncertainty on the spectral parameters of PSR\, J1357$-$6429, especially in the $\gamma$-ray band, makes it difficult to draw any conclusion.  As new data are taken by {\it Fermi},  it will be possible to better constrain the value of the pulsar's $\gamma$-ray photon index and verify the possible absence of spectral breaks and its dissimilarity with PSR\, J1048$-$5832.

\section{Summary and conclusions}

We used archival \vlt/\fors\ observations to perform the first deep optical investigations of the two  {\em Fermi} pulsars PSR\, J1357$-$6429 and  PSR\, J1048$-$5832.  We re-assessed the positions of the two pulsars from the analyses of all the available \chan\ observations and the comparison with the published radio positions. For PSR\, J1357$-$6429, this yielded a tentative proper motion $\mu = 0\farcs17 \pm 0\farcs055$ yr$^{-1}$  ($70^{\circ} \pm 15^{\circ}$ position angle), which needs to be confirmed by future radio-interferometry observations. For PSR\, J1048$-$5832, we concluded that its radio-timing position is either wrong or affected by large uncertainties due to timing irregularities. 
For  both pulsars, none of the objects detected around the \chan\  positions can be considered viable candidate counterparts on the basis of their relatively large optical flux, $\ga 3\sigma$ offset from the pulsar position,  and lack of peculiar colours with respect to the field stars.  

For PSR\, J1357$-$6429, we found a marginal evidence of a flux enhancement over the background at the edge of the \chan\ error circle, which might be due to the presence of a faint source ($I\approx 24.6$). However, the local crowding, with two relatively bright stars within a radius of $\sim 1\farcs5$ from the \chan\ position, and the lack of detections in the V and R-bands, make it problematic to determine whether such an enhancement is due to a background fluctuation, perhaps produced by the PSF wings of the two stars, or it is associated with a real source and hence to a putative pulsar counterpart.   Assuming that PSR\, J1357$-$6429 is not detected in the \vlt\ images, we determined $3 \sigma$ upper limit on its optical brightness of $V \sim 27$.  This implies an optical emission efficiency $\eta_{opt} \la 7\times 10^{-7}$, at  least a factor of 5 lower than the Crab pulsar and, possibly, more compatible with that of the Vela pulsar.
For PSR\, J1048$-$5832, 
our $3 \sigma$ upper limit of $V\sim 27.6$ implies an optical emission efficiency $\eta_{opt} \la 6 \times 10^{-6}$, which is still compatible with a Crab-like optical emission efficiency.
Finally, in both cases we did not find evidence for possible  optical counterparts of the PWNe detected in X-rays by \chan,  with the search complicated by the field crowding and the partial occultation at the pulsar position for  PSR\, J1357$-$6429, and by the presence of the molecular cloud complex for PSR\, J1048$-$5832. 

The \vlt\ observations presented here are close to the sensitivity limits achievable with 10m-class telescopes under sub-arcsec seeing conditions. In the case of PSR\, J1357$-$6429, deep, high spatial resolution images with the \hst\ are probably required to firmly claim an object detection from the flux enhancement seen at the \chan\ position. Moreover, if our tentative proper motion measurement is confirmed by future radio-interferometry observations, the pulsar will be close to occultation by the nearby Star B in mid 2012 and it will remain occulted for the following 10 years. Thus, prompt \hst\ follow-up observations in the I-band are the only way to achieve the optical identification of PSR\, J1357$-$6429. In the case of PSR\, J1048$-$5832,  the large interstellar extinction ($A_V \sim 5$) and the presence of the molecular cloud complex in the field hamper observations in the optical/near-ultraviolet. Deep observations in the near-infrared, either with the \hst\ or with adaptive optic device at 10m-class telescopes, might represent a better opportunity to spot a candidate counterpart to the pulsar.

\begin{acknowledgements}
The authors thank the {\em Fermi} Pulsar Timing Consortium, in particular David Smith, Ryan Shannon, and Simon Johnston, for checking for updated radio coordinates of the two pulsars. We also thank the {\em Fermi} Publication Board and the anonymous referee for their useful comments on the manuscript.
\end{acknowledgements}

\end{document}